\documentclass[11pt]{article}
\usepackage{moriond,epsfig}

\def\be{\begin{equation}}
\def\ee{\end{equation}}
\def\bea{\begin{eqnarray}}
\def\eea{\end{eqnarray}}

\begin{document}
\vspace*{4cm}
\title{HIGGS SEARCHES AT CMS AND ATLAS}

\author{ J. FERNANDEZ \\
on behalf of the ATLAS and CMS collaborations}

\address{Departamento de F\'{\i}sica, Facultad de Ciencias \\
C/ Calvo Sotelo s/n, 33007 Oviedo (Asturias), Spain}

\maketitle\abstracts{A summary of the sensitivity of the ATLAS and
CMS experiments at the LHC to discover a Standard Model Higgs
boson is presented.~Some prospects for Minimal Supersymmetric
Standard Model Higgs searches at LHC are also included.}

\section{Introduction}

The primary objective of the LHC is to elucidate the mechanism
responsible for electroweak symmetry breaking.  In the context of
the Standard Model (SM) the assumption of one doublet of scalar
fields gives rise to a scalar particle known as the Higgs
boson.~\cite{peter_higgs,AD} The Higgs mass is not predicted by
theory and, to date, direct experimental searches for the Higgs
have put a lower limit~\cite{lep} on its mass at $M_H >$ 114.4
GeV/c$^2$ and recently excluded~\cite{CDFD0} the $M_H \in
[160-170]$ range $\mbox{@}95\%$CL. A preferred value for the Higgs
mass, derived by fitting precision electroweak
data,~\cite{lepewwg} is currently $M_H = 90^{+36}_{-27}$ GeV/c$^2$
with an upper bound of 163 GeV/c$^2\mbox{ @ } 95\%$ CL. This bound
increases to 191 GeV/c$^2$ when the LEP-2 direct search limit of
114.4 GeV/c$^2$ is included in the fit.

Both the ATLAS and CMS experiments at the LHC, scheduled for
proton-proton collision data-taking  beginning Autumn 2009, have
been designed to search for the Higgs over a wide mass
range.~\cite{atlastdr,cmstdr} These proceedings summarize the
sensitivity for each experiment to discover a SM Higgs boson with
relatively low integrated luminosity per experiment ($1 - 5$
fb$^{-1}$) at $\sqrt{s}$ = 14 TeV, as well as recent developments
that have enhanced this sensitivity.  A brief discussion on the
sensitivity for these experiments to discover one or more of the
Higgs bosons from the minimal version of the supersymmetric
theories~\cite{mssm} (MSSM) is also included.

\section{Standard Model Higgs Production at the LHC}

\begin{figure}
\begin{minipage}[b]{0.45\linewidth}
\centering
\includegraphics[scale=1.25]{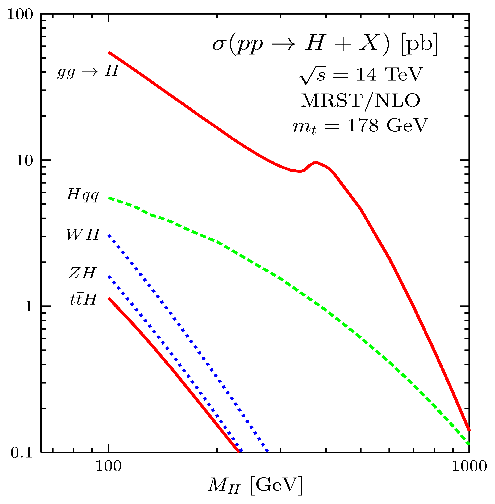}
\caption{Dominant Standard Model Higgs production cross-sections
at the LHC.$^2$} \label{fig:higgs_production}
\end{minipage}
\hspace{0.5cm}
\begin{minipage}[b]{0.45\linewidth}
\centering
\includegraphics[scale=1.28]{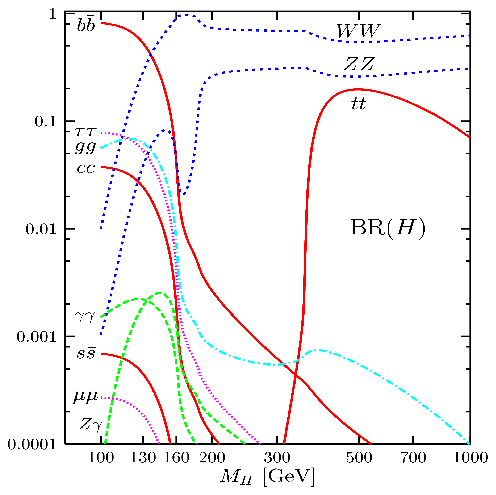}
\caption{Branching ratios for Standard Model Higgs decays.$^2$}
\label{fig:higgs_decay}
\end{minipage}
\end{figure}

The SM Higgs production NLO cross-sections at the LHC, as a
function of Higgs mass, are shown in
Figure~\ref{fig:higgs_production}. The dominant production
mechanism for SM Higgs boson production, which proceeds via a
top-quark loop, is the {\em Gluon-Gluon Fusion} mode
($gg\rightarrow H$) which suffers from large background at low
mass. The {\em Vector Boson Fusion} (VBF) process ($qq\rightarrow
Hqq$) is the second-most dominant production mode at the LHC
having a very distinct final state. {\em Associated Production}
modes, where the Higgs is produced via $q\overline{q}'\rightarrow
HW,$ $q\overline{q}\rightarrow HZ$ and
$gg,q\overline{q}\rightarrow t\overline{t}H$, have much smaller
cross-sections.  The presence of a $W, Z$ or top-quark alongside
the Higgs, or high-$p_T$ high-$\eta$ jets from VBF, allow for
triggering on events with Higgs in invisible final states.

\section{Higgs Discovery Final States}
The final states most suitable for discovery at the LHC vary
depending on the branching ratios (shown in
Figure~\ref{fig:higgs_decay}), which are a function of the Higgs
mass, and the relevant backgrounds.  For $M_H < 2M_W$ the dominant
decay mode is through $b\overline{b}$; however, due to the
enormous QCD background, the discovery is unlikely using this
channel. The $\gamma\gamma$ final state, which appears when the
Higgs decays via bottom, top and $W$ loops, has a small branching
fraction but an excellent $\gamma/\mbox{jet}$ separation and
$\gamma$ resolution help to make this a very significant channel.
$H\rightarrow\tau^+\tau^-$ is accessible through the VBF modes,
where the two struck quarks appear as high-$p_T$ jets in the very
forward (high-$\eta$) and opposite regions of the detectors.

If the Higgs mass is large enough to make the $WW$ and $ZZ$ modes
kinematically accessible, the $H\rightarrow WW^{(*)}$ final-states
are powerful over a very large mass range ($WW$ accounts for
$\sim$95\% of the branching ratio at $M_H \sim$160 GeV/c$^2$), as
is the $H\rightarrow ZZ^{(*)}\rightarrow 4l$ final state--the
latter of which is commonly referred to as the ``Golden Mode'' as
with four leptons in the final state the signal is easy to trigger
on and allows for full reconstruction of the Higgs mass.

Both ATLAS and CMS have conducted extensive fully-simulated {\tt
GEANT}-based~\cite{geant} Monte Carlo studies to determine the
experimental viability of all of these channels at NLO. A few of
these signatures are highlighted below; for details, the reader is
referred to their corresponding Technical Design
Reports~\cite{atlastdr,cmstdr} and updates.~\cite{PAS}

\subsection{$H\rightarrow\gamma\gamma$}
Despite the small branching ratio, $H\rightarrow\gamma\gamma$
remains a very attractive channel for $M_H <$ 140 GeV/c$^2$.
Genuine photon pairs from $q\overline{q}\rightarrow\gamma\gamma$,
$gg\rightarrow\gamma\gamma$ and quark bremsstrahlung comprise the
irreducible background, while jet-jet and $\gamma$-jet events,
where one or more jets are misidentified as photons, make up the
reducible background.  $Z\rightarrow ee$ events, with both
electrons misidentified as photons, can be reduced using
electron/photon separation techniques.  The sensitivity of this
channel is similar for both experiments; the high-granularity
liquid Argon calorimeter of ATLAS is capable of determining the
primary vertex with great precision, while CMS has a superior
energy resolution.  Studies conducted by both experiments consider
the signal and background to NLO. Both experiments have looked
beyond a simple cut-based analysis and enhanced the signal
significance by $\sim$35\%. With an integrated luminosity of 10
(30) fb$^{-1}$ the significance is just above $4\sigma$
($8\sigma$) for $M_H$ = 130 GeV/c$^2$, .

\subsection{$H\rightarrow ZZ^{(*)}\rightarrow 4l$ ($4e$, $4\mu$, and $2e2\mu$)}
At $M_H >$ 130 GeV/c$^2$, the 4-lepton channels gain in importance
on account of the precise energy reconstruction of both ATLAS and
CMS for electrons and muons. This channel has a very clean signal
with high branching ratio except at $M_H \sim 2 M_{W}$. Since an
excellent mass resolution ($1.5 - 2$ GeV/c$^{2}$ for $m_{H}$= 130
GeV/c$^{2}$) can be achieved, it is a powerful analysis in a wide
mass range. The dominant backgrounds for these channels are
$ZZ^{(*)}$, $Zb\overline{b}$ and $t\overline{t}$ production.
Through the use of impact parameter and lepton isolation
requirements the latter two can be significantly reduced.
Collectively, the significance for these channels is more than
$5\sigma$ for 30 fb$^{-1}$ of data in the whole mass range for
both experiments separately, which have comparable significance.

\begin{figure}
\begin{minipage}[htbp]{0.47\linewidth}
\centering
\includegraphics[scale=0.5]{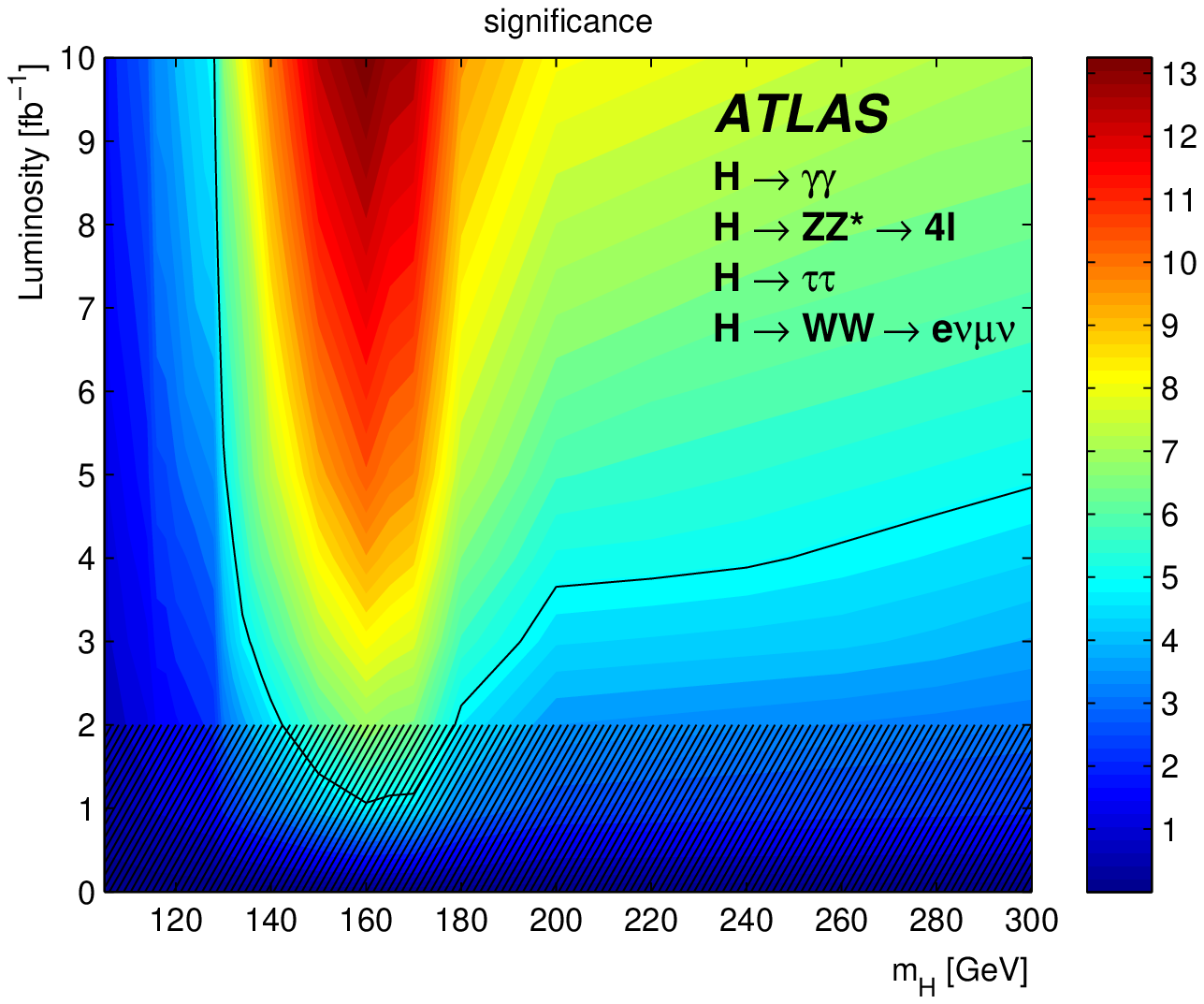}
\caption{The discovery potential at ATLAS$^6$ for Standard Model
Higgs boson searches. }
\label{fig:ATLASsm}
\end{minipage}
\hspace{0.5cm}
\begin{minipage}[htbp]{0.47\linewidth}
\centering
\includegraphics[scale=0.3]{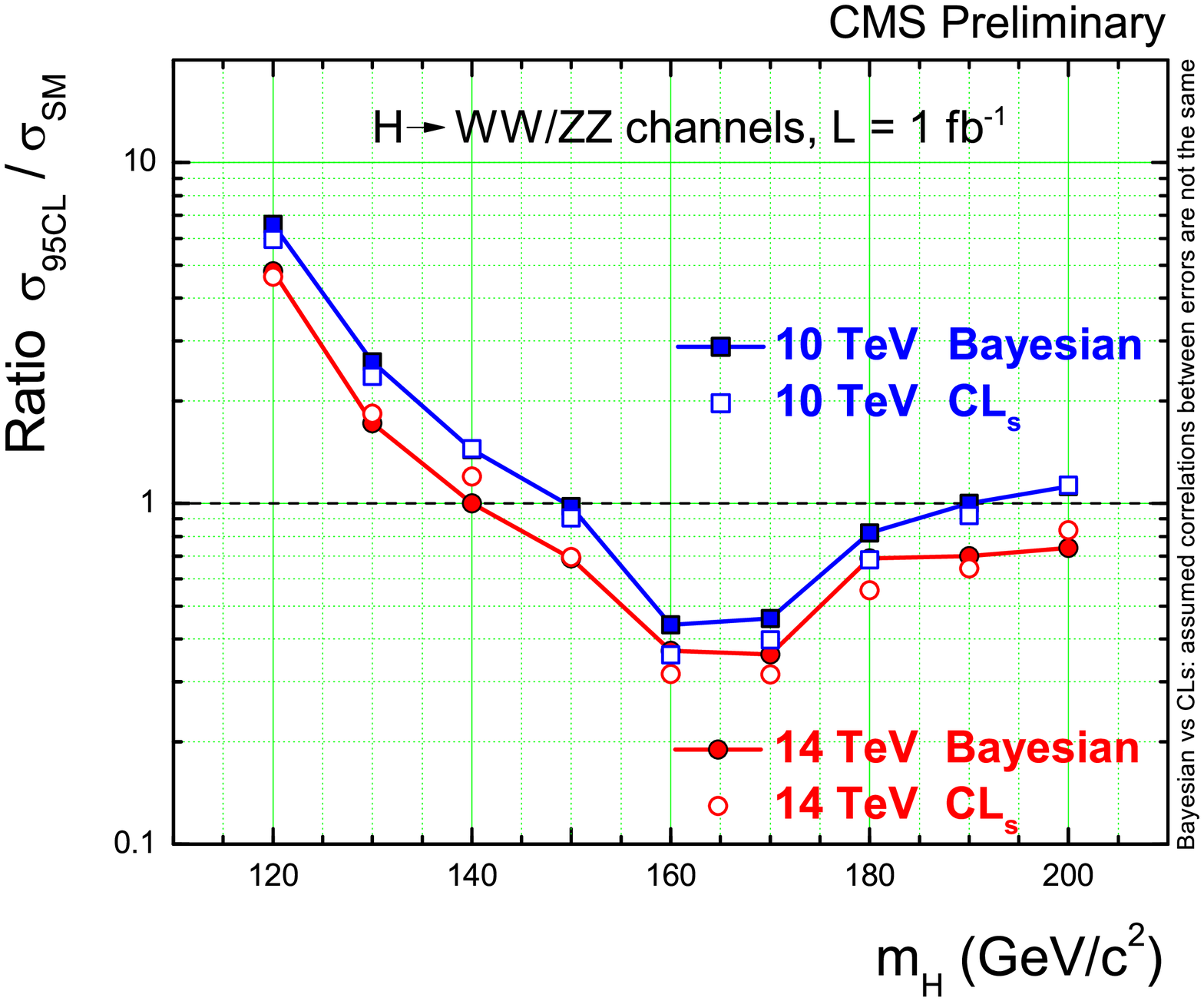}
\caption{The updated$^{10}$ discovery potential at CMS for
Standard Model Higgs boson searches.} \label{fig:CMSsm}
\end{minipage}
\end{figure}

\subsection{$H\rightarrow WW^{(*)}\rightarrow l\nu l\nu$ ($l=e$, $\mu$)}
As the branching ratio for a SM Higgs decaying to $WW$ is more
than $95\%$ at $\sim$160 GeV/c$^2$, this is the most significant
channel at that mass point.  Unlike other channels, in the
$H\rightarrow WW\rightarrow l\nu l\nu$ final state (ATLAS has
considered the $H+0j$ and $H+2j$ processes in the $e\nu \mu\nu$
channel only) full mass reconstruction is not possible and the
analysis is essentially reduced to a counting experiment;
therefore an accurate background estimate is critical. The
dominant backgrounds for this analysis are $WW$ and
$t\overline{t}$ production.  The former can be suppressed by
exploiting spin correlations between the two leptons while the
latter has been shown to be suppressed significantly by a jet
veto. Using NLO cross-sections and conservative estimates for the
effect of systematic uncertainties, a significance of around
$5\sigma$ for $M_H$ = 165 GeV/c$^2$ with an integrated luminosity
of $\sim$1 fb$^{-1}$ is estimated.

\subsection{$H\rightarrow \tau^+\tau^-$}
The distinct experimental signature of Higgs production via VBF
allows for search channels like $H\rightarrow \tau^+\tau^-$,
resulting in a very significant channel for low masses and
important for studying the coupling of Higgs to leptons. ATLAS and
CMS now both consider three final states, thus covering all
combinations of leptonically- and hadronically-decaying taus.
Triggering on the fully hadronic mode is currently under
investigation.  Despite the presence of multiple neutrinos in the
final-state, mass reconstruction can typically be done via the
collinear approximation where the tau decay daughters are assumed
to be in the same direction as their parent. A significance around
$5\sigma$ can be achieved with $\sim$ 30 fb$^{-1}$ at $M_H \sim$
115 GeV/c$^{2}$.

\section{Summary of Higgs Discovery Potential}
The expected significance combining various final states as a
function of SM Higgs mass and cumulated luminosity is summarized
in Figure~\ref{fig:ATLASsm} for the ATLAS experiment, while the
updated~\cite{PAS} combined and $\sqrt{s}$= 10 TeV projected
exclusion limit result for CMS in the two main search channels
($H\rightarrow ZZ^{(*)}$ and $H\rightarrow WW^{(*)}$) with 1
fb$^{-1}$ is shown on Figure~\ref{fig:CMSsm}. The discovery
potential at ATLAS and CMS is quite similar.

\begin{figure}
\begin{minipage}[htbp]{0.47\linewidth}
\centering
\includegraphics[scale=0.4]{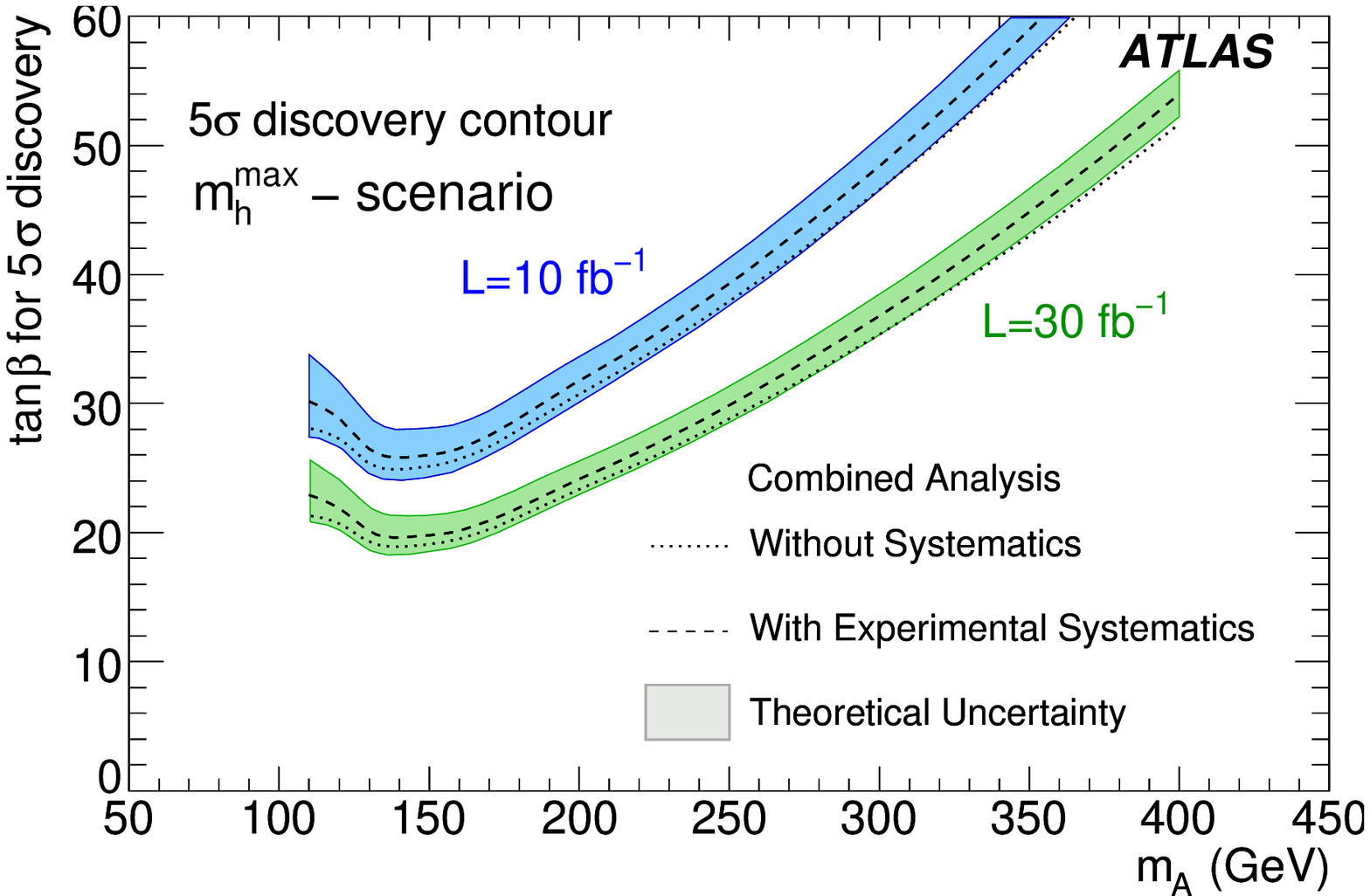}
\caption{ATLAS sensitivity$^6$ for the discovery of MSSM Higgs
bosons ($m_{h}^{max}$ scenario) in the $bbA/H/h \rightarrow
bb\mu\mu$ channel.} \label{fig:ATLASmssm}
\end{minipage}
\hspace{0.5cm}
\begin{minipage}[htbp]{0.47\linewidth}
\centering
\includegraphics[scale=0.28]{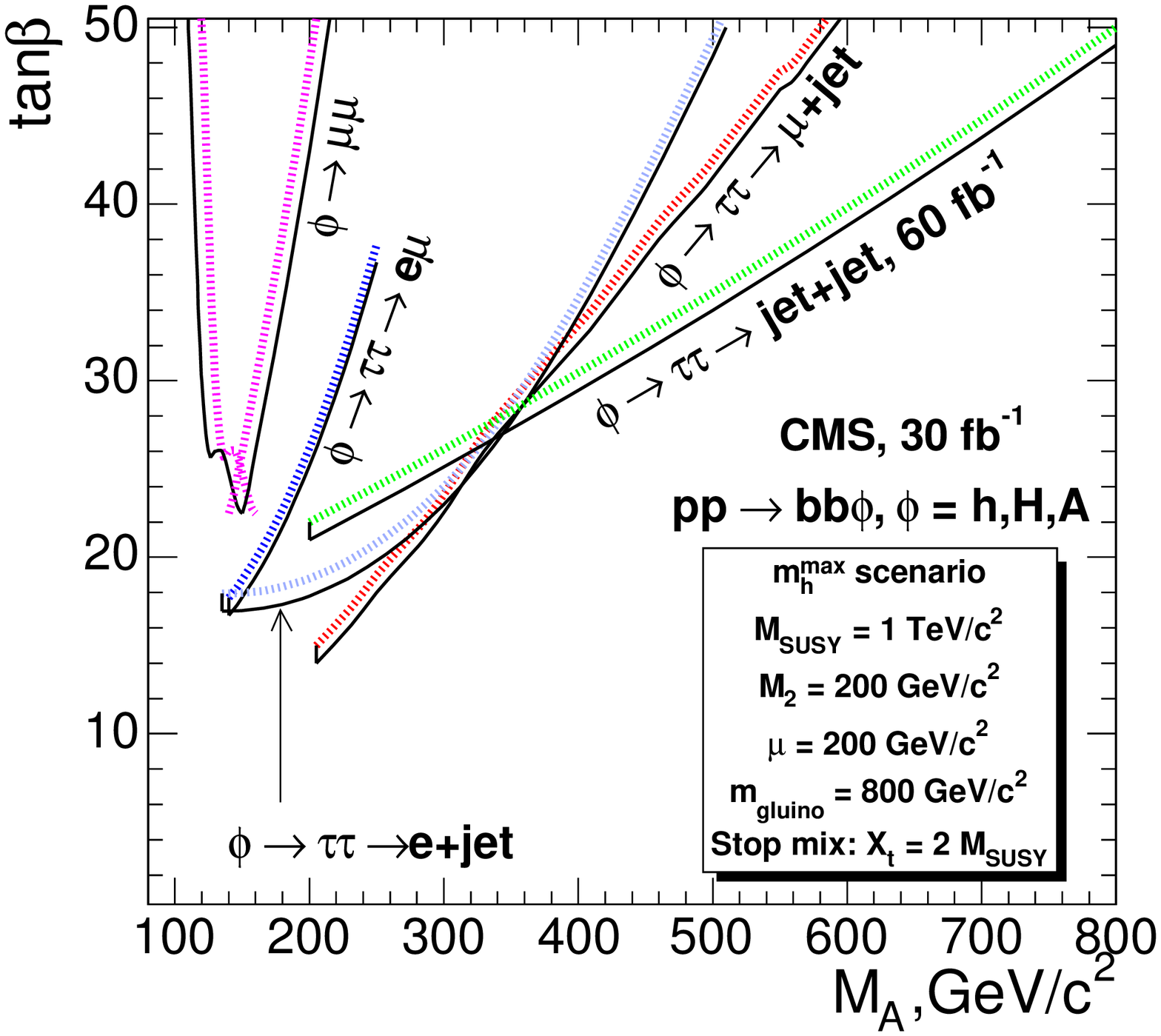}
\caption{The CMS$^7$ experiment $5\sigma$ discovery regions for
the neutral Higgs bosons $\phi$ $(\phi=h, H, A)$ produced in the
MSSM $m_{h}^{max}$ scenario.}

\label{fig:CMSmssm}
\end{minipage}
\end{figure}

Discovery prospects for the detection of MSSM Higgses ($A$, $h$,
$H$ and $H^{\pm}$) have also been
evaluated.~\cite{atlastdr,cmstdr} At tree-level, all Higgs masses
and couplings can be expressed in terms of $m_A$ and $\tan\beta$.
The complete region of the $m_A - \tan\beta$ parameter space
($m_A$ = 50 - 500 GeV/c$^2$ and $\tan\beta = 1 - 50$) should be
accessible to the LHC experiments. The sensitivity for the
discovery of MSSM Higgses, in the $m_{h}^{max}$ scenario is
summarized in Figures~\ref{fig:ATLASmssm} and~\ref{fig:CMSmssm}.
The $bbH/A$ production mode allows a reach extending down to $M_A
\sim$ 150 GeV/c$^{2}$, $\tan\beta \sim 15$ with the $\mu\mu$ and
$\tau\tau$ channels for 30 fb$^{-1}$ of data. SM Higgs searches
reinterpreted in the MSSM (mainly $h \rightarrow \gamma \gamma,
qqh/H \rightarrow qq\tau\tau$) cover the low and intermediate
$\tan \beta$ region.

\section*{Acknowledgments}
The author would like to thank the ATLAS and CMS Collaborations,
in particular A. Korytov, C. Mariotti, A. Nisati and Y. Sirois,
for their advice and support.

\end{document}